\documentclass[aps,showpacs]{revtex4}
\usepackage{epsfig}

\begin{document}

\title{Russian Doll Renormalization Group and 
Kosterlitz-Thouless Flows}
\author{Andr\'e  LeClair\footnote{On leave from 
 Newman Laboratory, Cornell University, Ithaca, NY
14853.},
Jos\'e  Mar\'{\i}a Rom\'an and  Germ\'an Sierra}
\affiliation{Instituto de F\'{\i}sica Te\'orica, UAM-CSIC, Madrid, Spain} 
\date{January 24, 2003}

\begin{abstract}
We investigate the previously proposed cyclic regime of
the Kosterlitz-Thouless renormalization group (RG) flows.  
The period of one cycle is computed in terms of the RG
invariant.  Using bosonization, we show that the theory
has $U_q (\hat{sl(2)})$ quantum affine symmetry, with $q$
{\it real}. Based on this symmetry, we  study two  possible 
S-matrices for the theory,  differing only by
overall scalar factors.   We argue that one S-matrix corresponds to
a continuum limit of the XXZ spin chain in the anti-ferromagnetic
domain $\Delta < -1$. The latter S-matrix has a periodicity in energy
consistent with the cyclicity of the RG. We conjecture that this S-matrix
describes the cyclic regime of the Kosterlitz-Thouless flows.
  The other S-matrix we investigate  is an analytic continuation
of the usual sine-Gordon one.  It  has an infinite number of resonances 
with masses that have a Russian doll scaling behavior that is also consistent with the period of the RG cycles computed from the beta-function.
Closure of the bootstrap for this S-matrix leads
to an infinite number of particles of higher spin with 
a mass formula suggestive of a string theory. 
\end{abstract}


\maketitle

\vskip 0.2cm

%
%
%
%
\def\oti{{\otimes}}
\def\lb{ \left[ }
\def\rb{ \right]  }
\def\tilde{\widetilde}
\def\bar{\overline}
\def\hat{\widehat}
\def\*{\star}
\def\[{\left[}
\def\]{\right]}
\def\({\left(}		\def\BL{\Bigr(}
\def\){\right)}		\def\BR{\Bigr)}
	\def\BBL{\lb}
	\def\BBR{\rb}
%
%
\def\zb{{\bar{z} }}
\def\zbar{{\bar{z} }}
\def\frac#1#2{{#1 \over #2}}
\def\inv#1{{1 \over #1}}
\def\half{{1 \over 2}}
\def\d{\partial}
\def\der#1{{\partial \over \partial #1}}
\def\dd#1#2{{\partial #1 \over \partial #2}}
\def\vev#1{\langle #1 \rangle}
\def\ket#1{ | #1 \rangle}
\def\rvac{\hbox{$\vert 0\rangle$}}
\def\lvac{\hbox{$\langle 0 \vert $}}
\def\2pi{\hbox{$2\pi i$}}
\def\e#1{{\rm e}^{^{\textstyle #1}}}
\def\grad#1{\,\nabla\!_{{#1}}\,}
\def\dsl{\raise.15ex\hbox{/}\kern-.57em\partial}
\def\Dsl{\,\raise.15ex\hbox{/}\mkern-.13.5mu D}
%
%
\def\th{\theta}		\def\Th{\Theta}
\def\ga{\gamma}		\def\Ga{\Gamma}
\def\be{\beta}
\def\al{\alpha}
\def\ep{\epsilon}
\def\vep{\varepsilon}
\def\la{\lambda}	\def\La{\Lambda}
\def\de{\delta}		\def\De{\Delta}
\def\om{\omega}		\def\Om{\Omega}
\def\sig{\sigma}	\def\Sig{\Sigma}
\def\vphi{\varphi}
%
%
\def\CA{{\cal A}}	\def\CB{{\cal B}}	\def\CC{{\cal C}}
\def\CD{{\cal D}}	\def\CE{{\cal E}}	\def\CF{{\cal F}}
\def\CG{{\cal G}}	\def\CH{{\cal H}}	\def\CI{{\cal J}}
\def\CJ{{\cal J}}	\def\CK{{\cal K}}	\def\CL{{\cal L}}
\def\CM{{\cal M}}	\def\CN{{\cal N}}	\def\CO{{\cal O}}
\def\CP{{\cal P}}	\def\CQ{{\cal Q}}	\def\CR{{\cal R}}
\def\CS{{\cal S}}	\def\CT{{\cal T}}	\def\CU{{\cal U}}
\def\CV{{\cal V}}	\def\CW{{\cal W}}	\def\CX{{\cal X}}
\def\CY{{\cal Y}}	\def\CZ{{\cal Z}}

\def\rvac{\hbox{$\vert 0\rangle$}}
\def\lvac{\hbox{$\langle 0 \vert $}}
\def\comm#1#2{ \BBL\ #1\ ,\ #2 \BBR }
\def\2pi{\hbox{$2\pi i$}}
\def\e#1{{\rm e}^{^{\textstyle #1}}}
\def\grad#1{\,\nabla\!_{{#1}}\,}
\def\dsl{\raise.15ex\hbox{/}\kern-.57em\partial}
\def\Dsl{\,\raise.15ex\hbox{/}\mkern-.13.5mu D}
%
%
%
\font\numbers=cmss12
\font\upright=cmu10 scaled\magstep1
\def\stroke{\vrule height8pt width0.4pt depth-0.1pt}
\def\topfleck{\vrule height8pt width0.5pt depth-5.9pt}
\def\botfleck{\vrule height2pt width0.5pt depth0.1pt}
\def\Zmath{\vcenter{\hbox{\numbers\rlap{\rlap{Z}\kern
0.8pt\topfleck}\kern 2.2pt
                   \rlap Z\kern 6pt\botfleck\kern 1pt}}}
\def\Qmath{\vcenter{\hbox{\upright\rlap{\rlap{Q}\kern
                   3.8pt\stroke}\phantom{Q}}}}
\def\Nmath{\vcenter{\hbox{\upright\rlap{I}\kern 1.7pt N}}}
\def\Cmath{\vcenter{\hbox{\upright\rlap{\rlap{C}\kern
                   3.8pt\stroke}\phantom{C}}}}
\def\Rmath{\vcenter{\hbox{\upright\rlap{I}\kern 1.7pt R}}}
\def\Z{\ifmmode\Zmath\else$\Zmath$\fi}
\def\Q{\ifmmode\Qmath\else$\Qmath$\fi}
\def\N{\ifmmode\Nmath\else$\Nmath$\fi}
\def\C{\ifmmode\Cmath\else$\Cmath$\fi}
\def\R{\ifmmode\Rmath\else$\Rmath$\fi}

\def\barray{\begin{eqnarray}}
\def\earray{\end{eqnarray}}
\def\beq{\begin{equation}}
\def\eeq{\end{equation}}

\def\gpar{g_\parallel}
\def\gperp{g_\perp}
\def\theta{h}

\def\Jb{\bar{J}}
\def\dx{\frac{d^2 x}{2\pi}}

\section{Introduction}

The Renormalization Group (RG) continues to be one of the most important
tools for studying the qualitative and quantitative properties of quantum
field theories and many-body problems in condensed matter physics.
A widely  important class of theories are those with RG fixed points in
the ultraviolet (UV) or infrared (IR), and our intuitive understanding
of the generic behavior of quantum field theory is largely based on
theories with these properties.  The notion that massive states
decouple in the flow toward the IR is an example of such an intuition. 
However fixed point behavior is not the only possibility, and physically
sound examples with other kinds of behavior are important to explore.  
This paper is concerned with {\it cyclic}  RG flows, 
whose possibility was considered as early as 1971 by Wilson\cite{Kwilson}.
However at the time no interesting models were known that exhibited
this behavior. 

Recently a cyclic RG behavior has been found in a number of models,
wherein the couplings return to their initial values after a 
{\it finite} RG time $\lambda$:
\beq
\label{Ii}
g(e^\lambda L ) = g(L),
\eeq
where $L$ is the RG length scale~\cite{nuclear,Wilson,BL,LRS}.
The models in~\cite{nuclear,Wilson} are problems in zero-dimensional
quantum mechanics.  The model in~\cite{LRS} is a natural extension 
of the BCS model of superconductivity and thus a many-body problem. 

The cyclic RG property eq.~(\ref{Ii}) has some important implications
for the spectrum of the hamiltonian.  Namely, if 
$\{ E_n , g, L \}$ is the spectrum of eigenvalues of 
the hamiltonian for a system of size $L$, then 
\beq
\label{Iii}
\{ E_n , g, e^\lambda L \} =  \{ E_n , g,  L \},
\eeq
i.e.\ the energy spectrum at fixed $g$ should reveal 
a periodicity as a function of $L$.  A nice feature of the
model in~\cite{LRS} is that in addition to the beta-function with
the cyclic property one could obtain analytic results for the spectrum 
using the standard BCS mean field treatment.  In this way one could study
the interplay between the cycles of the RG flow and the spectrum.  
The manner in which the spectrum reproduces itself after one RG cycle 
is dependent on the existence of an infinite number of eigenstates 
with the 'Russian doll' scaling behavior in an appropriate limit
\beq
\label{Iiii}
E_{n+1} \approx  e^\lambda E_n .
\eeq
In each cycle these eigenstates reshuffle themselves such that the
$(n+1)$'th state plays the same role as the $n$'th state of the previous
cycle.  

Another possible signature of a cyclic RG is a cyclicity of the 
S-matrix itself. 
  Applying  RG equations to the S-matrix $S(E)$, where
$E$ is the energy,   one expects:
\beq
\label{scyclicity}
S(e^{\lambda} E ) = S(E)
\eeq
The above equation  was anticipated in \cite{Kwilson}.
  
The models considered in~\cite{BL}  are relativistic
models of quantum field theory in $2d$.  What is surprising is that
this theory is in fact a well-known theory that arises in a multitude
of  physical
problems: an anisotropic 
 left-right current-current interaction that gives rise to
the Kosterlitz-Thouless (KT) flows.   The standard picture of these
flows, based on the one-loop approximation,  is shown in figure 1.  
The $\gpar$ axis is a line of IR or UV fixed points.  The flows
in the region $|\gperp| > |\gpar| $ on the other hand are already
peculiar in that they both originate and terminate at no clearly
identifiable fixed point.  In \cite{BL} it was proposed that this
region has a cyclic RG flow based on the all-orders beta-function 
conjectured in \cite{GLM}.  However since the flows extended beyond
the perturbative domain $|g_{\parallel, \perp}| < 1$, it remained
unclear whether other non-perturbative effects would spoil
the cyclicity.  Furthermore, at the time little was known about the
spectrum.    In any case, if  this proposal is correct it implies
that an important feature of the KT flows has been
overlooked.  

\begin{figure}[t!]
\begin{center}
\includegraphics[height=6 cm, angle=0]{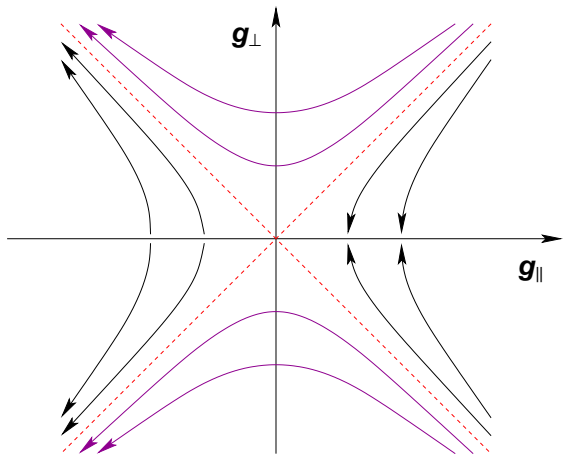}
\end{center}
\caption{Kosterlitz-Thouless flows.} 
\label{fig1}
\end{figure}

The main purpose of this article is to provide further evidence
for the cyclicity of the KT flows by proposing an S-matrix.  Let
us summarize our main results.  In theories with UV or IR fixed points
the universal properties encoded in the beta-functions are anomalous
dimensions of fields which are related to the slope of the beta-function
at the fixed point.  On the other hand, for a cyclic flow,  the only
universal, coordinate independent property is the period of one cycle. 
We compute the period $\lambda$ in section II in terms of the
RG invariant $Q \equiv  -\theta^2 /16$ 
 (at small coupling 
$ Q \approx \gpar^2 - \gperp^2 $), 
and find the result
$\lambda = 2\pi / \theta $.    

In section III we bosonize the currents and obtain a sine-Gordon
theory with a special form of the sine-Gordon coupling with both
real and imaginary parts. (See eq.~(\ref{bos4}).) Other regimes
of the KT flows are known to have an exact S-matrix description
corresponding to the sinh-Gordon and massive/massless sine-Gordon
theories.  In section IV we argue that the theory possesses
$\CU_q (\hat{sl(2)})$ symmetry with $q$ {\it real}. (The usual
sine-Gordon model corresponds to $q$ a phase.)   This symmetry
fixes the S-matrix up to an overall scalar factor. 
The only free parameter of this S-matrix
is $\theta$ which governs the period of the RG cycles,
and we find $q=-e^{-\pi \theta /2}$. 
 
For the simplest choice of the overall scalar factor,  our S-matrix 
can be understood as a certain relativistic continuum limit of the XXZ spin
chain with $\Delta < -1$.
In terms of $\Delta$,  the period of a cycle is
\beq
\label{0.2}
\lambda = \frac{\pi^2}{\cosh^{-1} (-\Delta) }.
\eeq
   In this case,  the signature of the
cylic RG is the periodicity of the S-matrix eq. (\ref{scyclicity}), 
where the period agrees precisely with the beta function computation 
in section II.  

The other possible overall scalar factor we investigate is
an analytic continuation of the usual sine-Gordon one.  The
resulting S-matrix is not real analytic, and for this reason
we believe the S-matrix related to the XXZ spin chain is the
correct one physically.  Nevertheless this second S-matrix has
some interesting properties we explore in the last section of the paper.  
 The soliton  S-matrix possesses poles
which  imply the existence of an infinite number of  resonances 
with mass
\beq
\label{Iv}
m_n = 2 M_s \cosh \frac{n\pi}{\theta} , ~~~~~~n= 1, 2, 3, ..., \infty,
\eeq
where $M_s$ is the mass of the soliton.  We show that these resonances
are actually  a spin 1 triplet and a singlet.
The above spectrum, eq.~(\ref{Iv}), which is rather novel in the subject
of integrable quantum field theory, is the Russian doll spectrum that
was anticipated based on the analogy with the results in~\cite{LRS}.
Namely, 
\beq
\label{Ivi}
m_n \sim M_s \, e^{n \pi/\theta} \quad\Longrightarrow \quad
m_{n+2} \approx e^\lambda m_n , ~~~~ n  \gg  \frac{\theta }{\pi },
\eeq
to be compared with eq. (\ref{Iiii}).   This property of the spectrum
is a strong indication of the cyclicity of the RG.
Furthermore,   
we find  that closing the S-matrix
bootstrap leads to resonances of higher spin
with a mass formula that suggests a string theory description. 
Theories with an infinite number of resonance poles were also
found in \cite{infres1,infres2}; however these theories 
appear to be rather different, the S-matrices being built out
of elliptic functions.

\section{Cyclic RG flows in anisotropic $SU(2)$ current perturbations}

We define the couplings $\gpar$, $\gperp$ as the following 
current-current perturbation of the $su(2)$ level $k=1$ Wess-Zumino-Witten
(WZW) model:
\beq
\label{1.1}
S = S_{WZW} + \int \dx  \( 4 \gperp (J^+ \Jb^- + J^- \Jb^+ ) - 4 \gpar J_3
\Jb_3 \),
\eeq
where $J^a$ ($\Jb^a$)  are the left (right) -moving currents, 
normalized as in~\cite{GLM}.

Let us first consider the well-known one-loop result. The beta functions
are 
\beq
\label{one1}
\frac{d\gpar}{dl} = - 4 \gperp^2 , \qquad \frac{d\gperp}{dl} 
= - 4 \gperp \gpar, 
\eeq
where the length scale is $L= a\, e^l$, with $a$ a microscopic
distance. We will refer to $l$ as the
RG `time'.    The flows possess the RG invariant:
\beq
\label{one2}
Q = \gpar^2 - \gperp^2, 
\eeq
so that the RG trajectories are hyperbolas.
We are interested
in the region $\gperp^2 > \gpar^2 $ where $Q$ is negative.  Let us
parametrize $Q$ as follows: 
\beq
\label{one3}
Q \equiv -\frac{\theta^2}{16} , \qquad \sqrt{Q} \equiv i \frac{\theta}{4},
\eeq
where $\theta > 0$.  Eliminating $\gperp$, one finds 
\beq
\label{one4}
\frac{d\gpar}{dl} = -4 \( \gpar^2 + \frac{\theta^2}{16} \).
\eeq
This is the same beta-function as in~\cite{LRS}.  The latter is
easily integrated:
\beq
\label{one5}
\gpar (l) = - \frac{\theta}{4} \tan \left(\theta (l - l_0 )\right),
\eeq
where $l_0$ is a constant.  One sees from eq.~(\ref{one5}) that 
$\gpar$ flows to $-\infty$ then jumps to $+\infty$ and begins a new
cycle. The periodicity of the RG flow is
$\gpar (l + \lambda_{\rm 1-loop} ) = \gpar (l) $ where
$\lambda_{\rm 1-loop} = \pi / \theta $.  

Though the 1-loop RG already indicates a cyclicity, since the flows
extend outside of the perturbative domain $|\gpar| < 1$ one cannot
conclude the flows are indeed cyclic based on the 1-loop approximation. 
However it was shown in~\cite{BL} that the cyclicity persists to 
higher orders in perturbation theory, as we now review.

The all-orders beta function proposed
in~\cite{GLM} is 
\beq
\frac{d \gpar}{dl} = \frac{-4 \gperp^2 (1+\gpar)^2 } {(1-\gperp^2 )^2 },
\qquad
\frac{d \gperp}{dl} = \frac{ -4 \gperp (\gpar + \gperp^2 ) }
{ (1-\gperp^2)(1-\gpar) }.
\label{1.2}
\eeq  
A few remarks on the status of the conjectured beta-function eq.~(\ref{1.2})
are in order.  A number of important checks were performed in~\cite{BL}.  
The most sensitive check was of the massless flows
that arise in the imaginary sine-Gordon theory defined by 
 $\gperp \to i \gperp$.  The above beta-function
correctly predicts the known non-perturbative relation between 
the anomalous dimensions in the UV and IR.  An all orders beta-function
was also proposed by Al.~Zamolodchikov~\cite{AlZ}, 
  and his result was quoted in~\cite{Lukyanov}.  
Zamolodchikov's argument was global in nature
and did not rely on summing up perturbation theory; the main input was
the known properties of the massless flows, thus his beta-function 
appears to be the unique one up to a change of coordinates that captures
the non-perturbative aspects of the flows.   It can be shown that
the beta-function in~\cite{Lukyanov} is equivalent to~(\ref{1.2}) 
under a change of coordinates.  Recently it has been argued that
there are additional contributions that first arise at 4-loops~\cite{ludwig}. 
Though this would seem to spoil the properties of the massless flows, 
arguments were given in~\cite{ludwig} that the new contributions would not. 
This is a rather technical issue and consequently we will assume the
above beta-function is correct.  Clearly, the present work
is a further check of the global properties of the conjectured
all-orders beta function.

The RG flows were analyzed in detail in~\cite{BL}.  It was shown there
that though the beta function possesses poles at~$\pm 1$,  the flows
approach the poles along non-singular trajectories and can be extended
to all length scales.  The analysis of the flows is simplified by 
recognizing that they possess a non-trivial RG invariant: 
\beq
\label{1.3}
Q = \frac{  \gpar^2- \gperp^2}{(1+\gpar)^2 (1-\gperp^2) }.
\eeq
(We have rescaled $Q$ by a factor of 16 in comparison to~\cite{BL}.) 

\def\sq{\sqrt{Q}}

One can use $Q$ to eliminate $\gperp$, obtaining 
\beq
\label{1.4}
\frac{d \gpar}{dl} = - 4 \frac{ \Bigl(1- (1+\gpar)^2 Q  \Bigr) 
\( \gpar^2 - (1+\gpar)^2 Q \) } {(1-\gpar )^2 }. 
\eeq
This is readily integrated:
\beq
\nonumber
l-l_0 = \inv{4 \sq} \[ \tanh^{-1} \( \frac{ \gpar (1-Q) - Q }{\sq} \) 
- \tanh^{-1} \( \sq (\gpar +1 ) \) \] + \log \( 
\frac{ \gpar^2 - (1+\gpar)^2 Q }{1 - (1+\gpar )^2 Q} 
\),
\label{1.5}
\eeq
where $l_0$ is a constant.  Note that there are no singularities 
at the location of the original poles of the beta function.

As described in~\cite{BL}, the RG flows possess a number of phases 
depending on the value of $Q$.  For the flows with UV or IR fixed points
the perturbing operators away from the fixed point are scaling fields,
i.e.\ their scaling dimension is constant along the flow.  Thus
for the flows with fixed points, the RG invariant $Q$ encodes 
these scaling dimensions.  The flows with fixed points all correspond
to $Q>0$ and are reviewed in more detail in the next section. 

The cyclic flows on the other hand correspond to $Q<0$.  Here, since
there are no UV or IR fixed points, $Q$ does not encode anomalous
dimensions, but rather the only universal feature of the flow,
namely the period of the cycles, which we now calculate.   
The coupling $\gpar$ flows toward $-\infty$, where it jumps to 
$+\infty$ and eventually returns to its initial value.  One
can see explicitly this jump from the solution as follows. Parameterizing
$Q$ as in eq.~(\ref{one3}), 
for $\theta$ large, and $|\gpar| \approx \infty $, the solution 
is 
\beq
\label{1.7}
\gpar \approx - \frac{4}{\theta} ~ \tan \[\frac{1}{2}\theta(l - l_0)\].  
\eeq
Thus when $l-l_0 = \pi / \theta $ the coupling jumps from 
$-\infty$ to $+\infty$.  

As before, 
define $\lambda$ as the period in $l$ for one cycle, eq.~(\ref{Ii}). 
Evidently
$\lambda = l_{-\infty} - l_\infty $ where $l_{\pm \infty}$ are
the RG times when $\gpar = \pm \infty$.  From the exact solution
eq.~(\ref{1.5}), and 
 the fact that $\tanh^{-1} (i\infty) = i\pi / 2 $,  
 one finds the simple result
\beq
\lambda = \frac{2 \pi}{\theta}. 
\eeq
Note that this is twice the 1-loop result. 

The RG results presented so far also give an indication
of the mass gaps of this theory. Indeed from eq.~(\ref{1.7})
we see that when $l -l_0 = \pi/\theta$ the coupling $g_\parallel$
becomes  $- \infty$, which implies a mass gap 
$M_0 \sim \frac{1}{a} e^{- \pi/\theta}$ associated to the
length scale $a \, e^{\pi/\theta}$ (we are assuming for simplicity
$ l_0 \ll \pi/\theta$ and $\theta \ll 1$). This mass is usually associated
with  the existence of two particles, the spinons, 
 in the application of  this model to the 
XXZ spin chain  
in the antiferromagnetic regime (see section V).

\def\bh{b}

\section{Bosonization}

The level-1 $su(2)$ current algebra can be bosonized as follows:
\beq
\label{bos1}
J^\pm = \inv{\sqrt{2}} e^{ \pm i \sqrt{2} \vphi } , \qquad
J_3 = \frac{i}{\sqrt{2}} \d_z \vphi,
\eeq
where $\vphi (z)$ is the $z=t+ix$-dependent part of a free massless
scalar field $\phi = \vphi (z) + \bar{\vphi} (\zb ) $.  Viewing
the $\gpar$ coupling as a perturbation of the kinetic energy term
and rescaling the field $\phi$ one obtains the action
\beq
\label{bos2}
S = \inv{4\pi} \int d^2 x \( \inv{2} (\d \phi )^2 +
\Lambda \cos \bh  \phi  \),
\eeq
where to lowest order $\Lambda \propto \gperp$. 
The field $\phi$ is normalized such that when $\Lambda = 0$, 
$\langle \phi (z,\zbar) \phi (0) \rangle = - \log z\zb $ 
and the scaling dimension of $\cos \bh \phi$ is $\bh^2$. 
(The coupling $\bh$ is related to the conventional sine-Gordon
coupling $\beta$ as $\bh^2 = \beta^2 / 4\pi $ so that the 
free fermion point corresponds to $\bh = 1$.)   It was shown
that the all-orders beta-function eq.~(\ref{1.2}) is consistent
with the known two-loop beta-function of the sine-Gordon theory
in~\cite{LecQH}.

\def\sQ{\sqrt{Q}}

For the flows with fixed points, i.e.\ $Q >0$, 
the coupling $\bh$ can be related to
$Q$ by properly matching the slope of the beta function at
the fixed points where $\gperp =0$.  The result is \cite{BL}
\beq
\label{bos3}
\bh^2 = \frac{2}{1+2\sqrt{Q}}  
\qquad \Longleftrightarrow \qquad
\inv{\sqrt{Q}} = \frac{2\bh^2}{2 - \bh^2}.
\eeq
The region $0 < \sqrt{Q} < \infty$ is the massive sine-Gordon phase
with $0<\bh^2 < 2$.  If $0 < \bh^2 <1$, i.e.\ $\sqrt{Q} > 1/2$,  
the spectrum contains bound states of the 
solitons and antisolitons called breathers,
which are absent if  $1 < \bh^2 <2$, i.e.\  $0 < \sqrt{Q} < 1/2$. 
When $\sQ <- 1/2 $, $\bh^2 < 0$, and $\bh $ is
purely imaginary, this corresponds  to the massive sinh-Gordon theory.
Finally when $-1/2  < \sQ < 0 $, the perturbation is irrelevant
since $\bh^2 > 2$ and this corresponds to the massless regime
with an IR fixed point. All these regions are plotted
in figure~2.

\begin{figure}[h]
\begin{center}
\includegraphics[height=6 cm, angle=0]{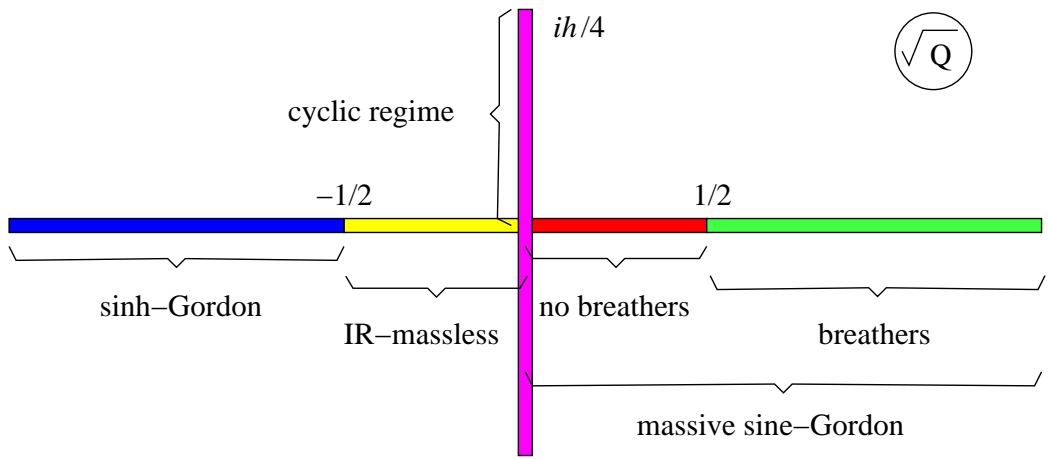}
\end{center}
\caption{Description of the different regimes of the
sine-Gordon model as a function of $\sqrt{Q}$. } 
\label{fig2}
\end{figure}

Parameterizing $Q$ as in eq.~(\ref{one3}), and making
an analytic extension of eq.~(\ref{bos3}), we see that 
the cyclic regime corresponds to 
\beq
\label{bos4}
\bh^2  = \frac{2}{ 1 + i \theta/2} = 
\frac{4}{(4 + \theta^2 )^{1/2}}\, e^{ -i \tan^{-1} (\theta/2 ) }.
\eeq
Since $\bh$ has both real and imaginary parts, this theory is neither
the sine-Gordon nor the sinh-Gordon theories. In this bosonized
description the hamiltonian does not appear to be hermitian 
however this is an artifact of the bosonized description since in
the current algebra description the action is real.  To lowest order,
the  origin
of the imaginary part of $b$ is the square-root in the rescaling
of the field involved in
going from eq.~(\ref{bos1}) to eq.~(\ref{bos2}).  Furthermore
the spin-chain realization in section VI also has a hermitian hamiltonian.

\section{Quantum Affine Symmetry and S-matrices}

In this section we
 will use some properties of the bosonized description to put forward
a conjecture concerning the spectrum and S-matrices.  
  The sine-Gordon theory is known to possess the 
$\CU_q (\hat{sl(2)} )$ quantum affine 
symmetry which commutes with the hamiltonian
and the S-matrix~\cite{nonlocal}, where 
\beq
\label{que}
q = e^{2\pi i / \bh^2 } = - e^{2\pi i \sQ }.
\eeq
The construction in~\cite{nonlocal} is valid regardless of value
of $\bh$. 
For the cyclic regime, though $\bh$ has real and imaginary parts, 
$q$ is real:
\beq
\label{que2}
q = - e^{ - \pi \theta /2 }.
\eeq
 The form of $\bh$ in eq.~(\ref{bos4}) is rather special
in that it leads to the real value of $q$ in eq.~(\ref{que2}); 
a more general complex $\bh$ would imply a complex $q$.

We point out that
the sine-Gordon theory is also known to possess another, dual, quantum
group structure $\CU_{\tilde{q}} (\hat{sl(2)} )$
which  encodes the commutation relations of
the monodromy matrix~\cite{Faddeev,Korepin}.
The dual quantum group is the algebraic structure underlying the
Bethe ansatz in the quantum inverse scattering method. 
The latter is closely tied to the hamiltonian itself, since the
trace of the monodromy matrix yields the integrals of motion.
Generally, if $q= e^{\pi a}$ then the dual is $\tilde{q} = e^{ \pi / a}$. 
Let us flip the sign of $q$, which in the field theory is
nothing more than a Klein transformation modifying the statistics
of the soliton fields.   Then 
the dual-$q$ is precisely the inverse of the finite RG transformation of one
cycle:  $\tilde{q}^{-1}  = e^{2\pi/ \theta}= e^\lambda$. 
Though this is an intriguing observation,  we will make no further use
of it in this paper.  

\def\rap{\beta}
\def\spec{\zeta}
\def\comb{\frac{\rap\theta}{2\pi} }
\def\Ga{\Gamma}

As for the usual sine-Gordon theory, we assume the theory contains
solitons of topological charge $\pm 1$.  In the XXZ spin chain, 
these solitons are the well-known spinons. (See the next section.) 
It was shown in~\cite{nonlocal} that the S-matrix is fixed up to 
an overall scalar factor by requiring it to commute with
the quantum affine  symmetry. 
We now describe the resulting soliton S-matrix in detail.  Let $\rap$
denote the usual rapidity that parameterizes the energy and momentum
of the soliton:
\beq
\label{5.1}
E = M_s \cosh \rap , ~~~~~~p = M_s \sinh \rap .
\eeq
Let $A_a (\rap )$ denote the creation operators for the solitons,
where $a=\pm$ denotes the topological charge.  The S-matrix
may be viewed as encoding the exchange relation of these operators:
\beq
\label{S1}
A_a (\rap_1 ) A_b (\rap_2) = S_{ab}^{cd} (\rap_1 -\rap_2 ) ~ 
A_d (\rap_2 ) A_c (\rap_1 ) .
\eeq
Requiring that the S-matrix commute with the quantum affine symmetry
leads to the following structure:
\beq
\label{5.3}
S (\rap ) =
\( \matrix{ S_{++}^{++} &0&0&0 \cr
0& S_{+-}^{+-} & S_{+-}^{-+} & 0 \cr 
0& S_{-+}^{+-} & S_{-+}^{-+} & 0 \cr
0&0&0 &S_{--}^{--} \cr 
} \) 
= - \frac{\rho (\rap )}{2i} 
\( \matrix{ q\spec - q^{-1}\spec^{-1}  &0&0&0 \cr
0& \spec - \spec^{-1}  & q-q^{-1} & 0 \cr  
0&q-q^{-1} & \spec - \spec^{-1} &0 \cr
0&0&0& q\spec - q^{-1}\spec^{-1}\cr
} \),
\eeq
where 
\beq
\label{5.3bis}
q = - e^{-\pi \theta /2 } , ~~~~~~~\spec = e^{-i\rap \theta / 2}.
\eeq
($\rap = \rap_1 - \rap_2 $.) 
The above S-matrix automatically satisfies the Yang-Baxter equation. 

The quantum affine symmetry does not constrain the overall scalar factor
$\rho$, however there are additional constraints coming from crossing
symmetry and unitarity.   
Crossing-symmetry requires:
\beq
\label{crossing}
S_{ab}^{cd} (\rap ) = S_{b\bar{c}}^{d\bar{a}} (i\pi - \rap ) 
,  \Longrightarrow  \rho (\spec ) = \rho (-\inv{q\spec} )
\eeq
Unitarity corresponds to $S^\dagger (\rap ) S(\rap ) = 1$. 
If the S-matrix is real analytic, i.e. 
\beq
\label{realanalytic}
S^\dagger (\rap ) = S(-\rap )
\eeq
then unitary requires
\beq
\label{unitarity}
S(-\rap ) S(\rap ) = 1
\eeq
The above equation leads  to the following constraint on the overall
scalar factor $\rho$:
\beq
\label{unit2}
\rho (\spec  ) \rho (\spec^{-1}  ) =
\frac{-4q^2 }{ (1-\spec^2 q^2 ) (1-\spec^{-2} q^2 ) }
\eeq

A minimal solution to the above equations 
is easily found iteratively. (See e.g. \cite{nonlocal}.)  One starts  with
a solution to eq. (\ref{unit2}),  
$\rho_0  = 2i q /( 1  - q^2 \spec^2 )$. This is made crossing symmetric 
as follows: 
 $ \rho_1 = \rho_0 ( \spec  ) \rho_0 ( -1/q\spec  )$.  This spoils unitarity, 
which can be corrected by multiplying by $\rho_1 (\spec^{-1} )$. 
Then one starts over again correcting crossing symmetry. The process
never terminates and leads to the infinite product:
\beq
\label{rhoxxz}
\rho (q, \spec ) 
= \frac{2i q}{1-q^2 \spec^2 } 
\prod_{n=0}^\infty 
\frac{ (1 - q^{4+4n} \spec^{-2} )(1-q^{2+4n} \spec^2 )}
{ (1-q^{4+4n} \spec^2 )(1-q^{2+4n} \spec^{-2} )}
\eeq
This  infinite product is convergent since $|q| < 1$.  

The above S-matrix is real analytic eq. (\ref{realanalytic}) 
and thus unitary in the usual sense.   Most remarkably, it
possesses a periodicity that is a clear signature of a cyclic
renormalization group.  Up to some minus signs, the 
S-matrix satisfies:
\beq
\label{Speriod}
S(\rap + \frac{2\pi}{\theta} ) = S(\rap ) 
\eeq
In the limit of large energies,  where $E\approx m e^\rap /2 $, 
the above equation implies 
eq. (\ref{scyclicity}).  
   
That the S-matrix presented in this section correctly describes
the current-current perturbation in the cyclic regime is the
main conjecture of this paper.   The lagrangian defines a hermitian
theory, and our S-matrix is unitary as it should be.  It also 
has a periodicity that is correctly predicted by the RG analysis of
section II.   In the next section we provide further support of
our conjecture based on the XXZ spin chain.

\section{Realization in the XXZ Heisenberg Chain}

In this section we give further support of the S-matrix proposed in 
the last section 
by taking a particular continuum limit of the XXZ spin chain. 
It is well-known that in the continuum limit the spin $1/2$ 
Heisenberg chain  is well described
by the  anisotropic current-current perturbation of  section II.
(See for instance~\cite{affleck,Lukyanov}.)

The spin  chain has the hamiltonian 
\beq
\label{1.9} 
H = - J
  \sum_{i=1}^N \( \sigma^x_i \sigma^x_{i+1} +  \sigma^y_i \sigma^y_{i+1}
 + \Delta ~  \sigma^z_i \sigma^z_{i+1} \).
\eeq
Let us first relate $\Delta$ to the couplings $\gpar$, $\gperp$ in the critical
(massless) regime, $-1 < \Delta < 1$.   In the current-current RG
description,  here the flows terminate in the IR along the critical
line $0<\gpar < 1$ with $\gperp = 0$. In this regime, 
$-1/2 <\sQ< 0$.  The spin-chain may be mapped onto a theory of
fermions via the Jordan-Wigner transformation. 
   Bosonizing these  Jordan-Wigner fermions, one
obtains the sine-Gordon theory.  
(See for instance~\cite{affleck,tsvelik,Lukyanov}.)  
In this  bosonized description, 
the flows arrive at the fixed point via the irrelevant 
operator $\cos ( \bh \phi ) $ with scaling dimension $\bh^2$.
The parameter $\bh$ is known
from the Bethe ansatz to be related to $\Delta$ as follows:
$\Delta = \cos 2\pi / \bh^2$.  In terms of $Q$:
\beq
\label{delta}
 \Delta = - \cos 2\pi \sq.
\eeq

Let us turn now to the cyclic 
regime,  where $-\infty < Q < 0$.  Let us assume that the formula~(\ref{delta})
is valid in this regime also.  Although $\sq$ is imaginary,
$\Delta$  is still real. 
Parameterizing $Q$ as in eq.~(\ref{one3}), $\sqrt{Q} = i \theta/4$, one finds 
\beq
\label{1.14}
\Delta = - \cosh {\pi\theta \over 2}. 
\eeq

 From eq.~(\ref{1.14}) one sees that the cyclic regime of the RG 
should correspond 
to the antiferromagnetic domain $\Delta < -1$ in the spin chain.
   In terms of $\Delta$,  the period $\lambda$
of the RG flows is given in eq.~(\ref{0.2}), namely
$\lambda = \pi^2/\cosh^{-1} (-\Delta) $.

There is one additional check of the above series of mappings. 
The spin chain is known to directly possess the $\CU_q (\hat{sl(2)})$
symmetry on the lattice~\cite{MJ}, where $q$ is related to 
$\Delta$ as follows:  
\beq
\label{1.15}
\Delta = \inv{2} (q + q^{-1} ). 
\eeq
Observe that the above equation is consistent with eqs.~(\ref{1.14})
and (\ref{que2}).

\def\am{\rm am}
\def\dn{\rm dn}

The XXZ spin chain with $\Delta < -1$ was studied in \cite{MJ}
using the underlying $\CU_q ( \hat{sl(2)} )$ symmetry on the lattice. 
One feature of this algebraic approach is the construction of
so-called type II vertex operators $Z_a$, $a=\pm 1$, which
can be interpreted as creation operators for the fundamental
spinons.  These operators satisfy the exchange relation
\beq
\label{ap1}
Z_a (\spec_1 ) Z_b (\spec_2 ) = R_{ab}^{cd} (\spec_1 / \spec_2 ) ~
Z_d (\spec_2 ) Z_c (\spec_1 ).
\eeq
(Compare with eq.~(\ref{S1})). 
The matrix $R$ in the above equation can be interpreted as
the S-matrix for the spinons on the lattice.  An S-matrix for
this regime of the spin chain was also obtained from the Bethe
ansatz in~\cite{Nep}.  The explicit form
of $R$ is precisely as in eq.~(\ref{5.3}) where now $q$ parametrizes
$\Delta$ as in eq. (\ref{1.15}) and is thus the same as in eq.~(\ref{que2}).  
The expression for $\rho$ on the lattice is the same as in
eq. (\ref{rhoxxz}); in the algebraic construction of\cite{MJ}, 
the S-matrix is just the universal R-matrix for
the quantum affine algebra.

On the lattice,  the particles have energies expressed in terms of
elliptic functions.  Letting
\beq
\label{ap4}
\spec = -i\, e^{i \alpha}, 
\eeq
then in terms of $\alpha$ the energy and momentum are 
\beq
\label{ap5}
E(\alpha ) = \frac{2K}{\pi} \sinh \frac{\pi K'}{K} \dn \( 
\frac{ 2 K \alpha} { \pi}  \) , 
~~~~~
p(\alpha ) = \am \(  \frac{2K\alpha}{\pi} \) - {\pi \over  2}.
\eeq
(See~\cite{Baxter,elliptic} for definitions.)  In the above formulas the nome
of the elliptic functions is $|q| = e^{ -\pi \theta /2 } = 
e^{ -\pi K'/ K } $ which
is between $0$ and $1$.  

Let us take a particular continuum limit which amounts to 
letting  the momentum and also $\theta$ to be small. 
Let us first take the momentum to be small.  From the definition of
$\am(u)$, one finds that $p(\alpha = \pi /2 ) = 0$.  Letting
$\tilde{\alpha} = \alpha - \pi/2 $, from the definition of $\am(u)$ 
one has 
\beq
\label{ap6}
K + \frac{2K\tilde{\alpha}}{\pi} = \int_0^{p + \pi /2 } 
\frac{d \psi}{\sqrt{1-k^2 \sin^2 \psi }},
\eeq
which can be transformed to 
\beq
\label{ap7}
\frac{2K\tilde{\alpha}}{\pi} = \int_0^{p } 
\frac{d \psi'}{\sqrt{k'^2 + k^2 \sin^2 \psi' }}.
\eeq
Above, $k$, $k'$ are the standard moduli of the elliptic functions. 
When $p$ is small, $\sin \psi' $ can be approximated by $\psi'$ and one
finds
\beq
\label{ap8}
p = \frac{k'}{k}  \sinh \( \frac{ 2 K k \tilde{\alpha}}{ \pi} \).   
\eeq

We now let $|q|$ approach one, i.e.\ $\theta$ small but not zero. 
Using known expressions in terms of the dual nome
$\tilde{q} = e^{ -2\pi / \theta } = e^{-\lambda}$, which goes to zero in the 
limit~\cite{elliptic}, one has 
\beq
\label{ap9} 
K \approx \frac{\pi}{\theta} \( 1 + 4 \tilde{q} \) ,
\qquad K' \approx \frac{\pi}{2} \( 1 + 4 \tilde{q} \),   
\qquad
k \approx 1-8 \tilde{q}, ~~~~~~k'\approx 4 \sqrt{\tilde{q}}.
\eeq
Thus in this limit one finds
\beq
\label{ap10}
p \approx  - k' \sinh \rap, 
\eeq
where the rapidity $\rap$ is identified as
\beq
\label{rapid}
\rap = - \frac{2\tilde{\alpha}}{\theta}.
\eeq

Turning to the energy, we need the identity
\beq
\label{ap11}
\dn (u ) = \sqrt{ 1- k^2 \sin^2 (\am(u)) }.
\eeq
Using the limits eq.~(\ref{ap9}) one finds 
\beq
\label{ap12}
E = \pi k' \cosh \rap.
\eeq
Thus the spinon velocity is $\pi$ and the mass goes with 
$k'=4 \sqrt{\tilde{q}}
= 4 e^{- \pi /\theta}$. The latter result agrees with the RG estimate
of the mass gap $M_0 \sim \frac{1}{a} e^{- \pi/\theta}$ made
in section II. Identifying the rapidity  $\rap$ from eq.~(\ref{ap10}) one
finds that the spectral parameter $\spec = e^{-i\rap \theta / 2 }$,
which agrees with eq.~(\ref{5.3bis}).

To summarize this section,  we have shown that a particular
continuum limit of the XXZ spin chain exists which leads to the 
 S-matrix conjectured in the last section.     
We also emphasize that in taking the  particular continuum limit described
above, 
we  have implicitly provided an ultra-violet completion 
of  the XXZ spin chain by defining a limit where the dispersion
relation is relativistic.

\section{Analytic continuation of the sine-Gordon S-matrix}

In this section we investigate another solution to the constraints
in section IV on the S-matrix.  Based on the bosonized description 
of section III, it is natural to consider the analytic extension of
the sine-Gordon S-matrix to the complex value of the coupling
$b$ given in eq. (\ref{bos4}).  We will refer to the resulting
model as the cyclic sine-Gordon model.  
Since the usual sine-Gordon S-matrix also commutes with the quantum
affine symmetry, the only difference between the cyclic sine-Gordon
 S-matrix
and the one proposed in section IV is in the overall scalar factor
$\rho$.     Extending the formulas in~\cite{ZZ} to the regime~(\ref{bos4}),
one finds the S-matrix is given by eq. (\ref{5.3}) but with 
$\rho $ given by $\rho_{SG}$, where:
\beq
\label{5.4}
\rho (\rap )_{SG}  = - \inv{ \pi  } \Gamma\( \kappa  \) 
\Ga \( 1- z \) \Ga \( 1 - \kappa  + z \) 
\prod_{n=1}^\infty 
\frac{ F_n (\rap ) F_n (i\pi - \rap )}{F_n (0) F_n (i\pi ) }, 
\eeq
with 
\beq
\label{5.5}
F_n (\rap ) = \frac
{ \Ga \( 2n\kappa  - z \) \Ga \( 1 + 2n\kappa  -z \) }
{\Ga \( (2n+1)\kappa  - z \) \Ga \( 1 + (2n-1) \kappa -z\) }. 
\eeq
where  we have defined
\beq
\label{defs}
\kappa = \frac{i\theta}{2} , ~~~~~z = -i \frac{\beta \kappa}{\pi} = 
\frac{\beta\theta}{2\pi}.
\eeq
The standard S-matrix for the massive sine-Gordon model
is given by the same formulas~(\ref{5.3},\ref{5.4},\ref{5.5})
with the following substitutions:
\beq
\kappa = \frac{8 \pi}{\gamma}, \qquad 
z = - \frac{8 \beta i}{\gamma} , \qquad
q= - e^{8 \pi^2 i /\gamma}, \qquad
\spec  = e^{- 8 \pi \beta/\gamma},
\label{5.6}
\eeq
where $\gamma$ is related to $b$ by the equation
\beq
\frac{\gamma}{8 \pi} = \frac{b^2}{2 - b^2} = \frac{1}{2 \sqrt{Q}}
,\qquad \sqrt{Q} > 0.
\label{5.7}
\eeq
Hence the relationship between the massive sine-Gordon S-matrix
and the  S-matrix considered in this section  amounts simply to the analytic
continuation
\beq
\theta 
\quad \longleftrightarrow \quad 
- \frac{16 \pi i}{\gamma}.
\label{5.8}
\eeq

The above S-matrix satisfies  the constraints of section
IV, including algebraic unitarity eq. (\ref{unitarity}), however
unlike the S-matrix in section IV it is not real analytic 
eq. (\ref{realanalytic}), and thus  cannot be considered unitary
in the usual sense.  As discussed in \cite{Mira}, in theories
where parity is broken, real analyticity can be violated, 
and replaced by hermitian analyticity. For simplicity
consider a diagonal scattering theory where the 
S-matrix for the scattering of particles $A_{a_1} (\beta_1 )$ 
and   
 $A_{a_2} (\beta_2 )$ is given by $S_{a_1 a_2} (\beta_1 - \beta_2 )$.
For a parity non-invariant theory, 
$S_{a_1 a_2} (\beta_1 - \beta_2 ) \neq 
S_{a_2 a_1} (\beta_1 - \beta_2 )$.
Unitarity, $S^\dagger S = 1$, requires
\beq
\label{hermana}
S^*_{a_2 a_1 } (\rap_1 - \rap_2 ) 
S_{a_1 a_2 } (\rap_1 - \rap_2 ) = 1 
\eeq
Imposing the requirement of hermitian analyticity
\beq
\label{hermana2}
S_{a_2 a_1} (\rap_1 - \rap_2 ) = S^*_{a_1 a_2 }(\rap_2 -\rap_1)
\eeq
the constrain of unitarity becomes
\beq
\label{hermana3}
S_{a_1 a_2} (\rap ) S_{a_1 a_2} (-\rap) = 1
\eeq
Note that the latter corresponds to algebraic unitarity
eq. (\ref{unitarity}) which is satisfied by the cyclic
sine-Gordon S-matrix.    We will henceforth assume the lack
of real analyticity can be understood in this way.

The analytic extension eq. (\ref{bos4}) is reminiscent of
the staircase models\cite{staircase1,staircase2}, however
there are some important differences.  The staircase model
corresponds to $b^2 = -2 ({1/2 +i\theta_0/2\pi})/({1/2 - i\theta_0/2\pi})$.
In terms of $\sqrt{Q}$, it corresponds to 
$\sqrt{Q} = -1/(1+i\theta_0/\pi)$.  For small $\theta_0$, 
this is a deformation of $\sqrt{Q} = -1$, which is in the sinh-Gordon
regime.  The S-matrix is thus an analytic extension of the
{\bf sinh}-Gordon S-matrix, with a single resonance, 
 whereas ours is a deformation of
the sine-Gordon one.   Furthermore, for the staircase model,
the c-function, though it roams, monotomically decreases.
For a model with a cyclic RG we expect a cyclic c-function
(see the conclusion).

\subsection{Resonances}

The  S-matrix  for the cyclic sine-Gordon model 
has numerous poles which can be interpreted
as resonances.   Interestingly,  other known  models 
that violate parity 
real analyticity also have resonances ~\cite{Mira}.
  More generally,  consider a pole at
$\beta = \mu - i \eta  $ with $\mu,~\eta > 0$,  in the S-matrix 
for the scattering of two particles of masses $m_1$, $m_2 $. As
discussed in~\cite{braz,mira2,Fring} this corresponds to a resonance
of mass $M$ and inverse lifetime $\Gamma$ where
\beq
\label{res1}
\(M- \frac{i \Gamma}{2} \)^2 = m_1^2 + m_2^2 + 2m_1 m_2 \cosh (\mu -i\eta ). 
\eeq
Equivalently:
\barray
\nonumber
M^2 - \frac{\Gamma^2}{4} &=& m_1^2 + m_2^2 + 2 m_1 m_2 \cosh \mu \cos \eta,
\\ 
\label{res2}
M\Gamma &=&  2 m_1 m_2 \sinh \mu \sin\eta.
\earray

The soliton  S-matrix has an infinite number of 
poles at $\beta_n = 2\pi n /\theta$
and $\bar{\beta}_n  = i\pi - 2\pi n / \theta $ with $n$ a positive
integer. These poles are absent in the S-matrix proposed in
section IV.  The poles at $\beta_n$ yield a real mass and correspond
to resonances with
an infinitely long lifetime $\tau = 1/\Gamma = \infty$, 
while the poles at 
$\bar{\beta}_n$ yield an imaginary mass and are not physical
(see table 1). 
It is interesting to compare this situation to what happens
for the usual massive
regime of the sine-Gordon theory where $0<b^2 < 2$. In this
case, when $\gamma < 8 \pi$,
the poles at  $\bar{\beta}_n$ lead
to the breather bound states, while those at 
$\beta_n$ are these same poles in the crossed
channel.

\begin{center}
\begin{tabular}{|c|c|}
\hline 
Cyclic sine-Gordon & Massive sine-Gordon \\
\hline \hline 
{\bf Resonances} &  Breathers in crossed channel \\
$\beta_n = \frac{ 2 \pi n}{\theta}$ 
& $\beta_n = \frac{i n \gamma}{8}$ \\
$m_n = 2 M_s \cosh \frac{n \pi}{\theta} \; \;  
(n=1, \dots, \infty)$ & \\
\hline
Resonances in crossed channel & {\bf Breathers}\\
$\bar{\beta}_n = i \pi - \frac{ 2 \pi n}{\theta}$ 
& $\bar{\beta}_n = i (\pi- \frac{ n \gamma}{8})$ \\
& 
$m_n = 2 M_s \sin \frac{n \gamma}{16}  \; \; 
(n=1, \dots < \frac{8 \pi}{\gamma})$ \\
\hline

\end{tabular}

\vspace{0.5 cm}
Table 1.- 
Poles and masses of the cyclic and the massive
sine-Gordon models.
\end{center}

Since the poles $\beta_n$ are right on the cut in the $s$-plane
($s= (p_1 + p_2)^2$),   
it is desirable  to incorporate a small imaginary part $-i\eta$ taking
them off the cut and giving the particles a large finite lifetime.
This can be viewed as a kind of ``$i\ep$'' prescription as follows.
Letting
\beq
\label{res4}
\theta \to \theta + i \ep,
\eeq
with $\ep$ very small,  
the poles are now at 
\beq
\label{res5}
\beta_n \approx \frac{2\pi n}{\theta} - i \eta_n , ~~~~~~
\eta_n \approx  \ep  \frac{2\pi n }{\theta^2}. 
\eeq
The prescription eq.~(\ref{res4}) does not spoil crossing-symmetry,
algebraic unitarity, nor the Yang-Baxter equation. 
As $\ep  \to 0$, from eq.~(\ref{res2}) 
the mass $m_n$ and width $\Gamma_n$ of the 
n-th resonance is 
\beq
\label{res6}
m_n \approx 2 M_s \cosh \frac{n\pi}{\theta} , ~~~~~~
\Gamma_n \approx \ep \, \frac{4\pi n}{\theta^2} M_s \sinh \frac{n\pi}{\theta}. 
\eeq

Each resonance of mass $m_n$, $n=1,2,..\infty$, actually corresponds
to four particles transforming in the direct sum of the q-deformed
spin 1 and singlet representations. One can easily
see this by noting that the entries $S_{++}^{++}, S_{+-}^{-+}$
and their charge conjugates  are non-zero at the poles $\beta_n$,
while  $S_{+-}^{+-}$ and  $S_{-+}^{-+}$ are zero. 
If the fundamental solitons have
topological charge $\pm 1$, then each resonance is a triplet 
with topological charges $(-2, 0, 2)$ or a singlet with
topological charge $0$. 
This is in contrast to the
usual breather bound state poles at $\bar{\beta}_n $ which occur
in the scattering channel of a soliton with an anti-soliton and are
thus singlets.   
($S_{++}^{++}$ and $S_{--}^{--}$ are zero  at the pole $\bar{\beta}_n $).

In the language of the fusion procedure~\cite{fusion}, which
is equivalent to the bootstrap,   the poles 
$\bar{\beta}_n$ project onto the singlet whereas $\beta_n$ 
projects onto the ($q$-deformed)  
spin-1 and spin-0 representations. 
This can be seen explicitly by going to the so-called homogeneous
gradation.  
The soliton S-matrix can be expressed as
\beq
\label{fus1}
S (\spec, q) = - \frac{\rho}{2i} \, 
\sigma_{21} \, \tilde{R} \, \sigma_{12}^{-1} \, P ,
\eeq
where 
$\sigma_{12} = \spec_1^{H/2} \otimes \spec_2^{H/2}$, 
$\sigma_{21} = \spec_2^{H/2} \otimes \spec_1^{H/2}$,
with $H = {\rm diag}(1,-1)$ the topological charge and 
$P$ the permutation operator:
\beq
\label{fus2}
P= \( \matrix{  1&0&0&0\cr
                0&0&1&0\cr
                0&1&0&0\cr
                0&0&0&1\cr 
             }  \).
\eeq
The matrix $\tilde{R}$ can then be expressed in terms of
the projectors $P_0$ and $P_1$  onto the q-deformed spin $0$
and spin $1$ representations:
\beq
\label{fus3}
\tilde{R} (\spec ) = 
(\spec q - \spec^{-1} q^{-1} )P_1 + (\spec^{-1} q - \spec q^{-1} ) P_0, 
\eeq
where 
\beq
\label{fus4}
P_0 = \inv{1+q^2}  \( \matrix{
            0&0&0&0\cr 
           0&1&-q&0 \cr 
            0&-q&q^2&0\cr
            0&0&0&0\cr } \),
\eeq
and $P_1 = 1-P_0$.   
The usual sine-Gordon breather poles at $\bar{\beta}_n$ correspond 
to $\spec = (-)^{n+1} q^{-1}$ and one sees that at this
$\spec$,  $\tilde{R} \propto P_0$, showing that the breathers
are indeed singlets.  A pole at $\spec = q$ projects onto the
irreducible spin 1 representation.  However our resonance poles
at $\beta_n $ correspond to $\spec = (-)^n$ where
$\tilde{R} \propto P_0 + P_1 $.   
The resonances, as the sine-Gordon breathers~\cite{ZZ}, also carry a
C-parity given by $(-1)^n$ for the triplets and $(-1)^{n+1}$
for the singlets.    


The above spectrum of masses has scaling properties that are
consistent with a cyclic RG.  As argued in section III 
the spinon mass $M_s $  must be proportional
to $1/L$, where $L$ is the system size.  
In the limit $n\gg \theta / \pi$ or $\theta \ll 1$,
eq.~(\ref{res6}) becomes, 
\beq
m_n(L) \sim \frac{1}{L} \, e^{n \pi/\theta}, \qquad 
n\gg \theta / \pi,
\label{res6bis}
\eeq
which exhibits the Russian doll scaling property
\beq
\label{res7}
 m_n(e^{-\lambda}  L ) \approx   m_{n+2} (L),
\eeq
similar to what was found in~\cite{LRS}.  Thus the way the spectrum
reproduces itself after one RG cycle is that $m_{n}$ at a given
length $e^{-\lambda}L$ plays the same role as $m_{n+2}$ 
at the longer length $L$ (see figure~3). Notice 
that the jump by two in $n$ is consistent with the $C-$parity
of the resonances. 
Eq.~(\ref{res7}) also shows that after one RG cycle
two new low energy masses appear in the spectrum.
This gain of the
lowest energy states in the spectrum is what allows
for the reshuffling of resonances
after one cycle and seems to be an essential ingredient
of any cyclic RG.

\begin{figure}[h]
\begin{center}
\includegraphics[height=6 cm, angle=0]{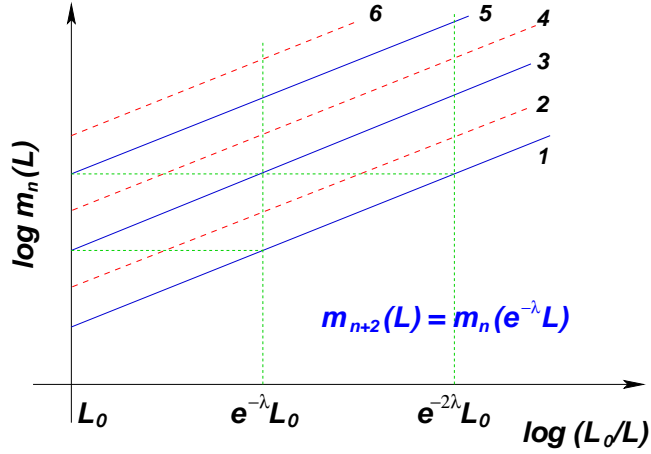}
\end{center}
\caption{Log-log plot of the resonance masses
$m_n(L)$ versus $1/L$, where $L$ is the length scale
of the system. $L_0$ denotes the maximal size of
the system. The Russian doll scaling of eq.~(\ref{res7})
is described by the dotted lines. 
} 
\label{fig3}
\end{figure}

In a finite size system we expect to have a finite number
of resonances,  $n_{\rm res}$,  which can be estimated as follows.
If in   every RG cycle we get two more resonances, then 
after $p$ cycles with $L \sim a \, e^{p \lambda}$,
we expect to have $n_{\rm res} = 2 p$ resonances,
with 
\beq
\label{res8}
n_{\rm res} \sim \frac{\theta}{\pi} \log \frac{L}{a}. 
\eeq
This kind of equation was checked numerically for the Russian
Doll BCS model\cite{LRS}.

\subsection{Closing the bootstrap: stringy spectrum}

In the limit that the regulator $\ep$ goes to zero, the resonances
are stable particles and can exist as asymptotic states. 
Thus, one should attempt to  close the bootstrap including
the resonances in asymptotic states as proposed in~\cite{Fring}. 
Since the bootstrap is equivalent to the fusion construction 
in the theory of quantum affine algebras,  it is clear for example
that the resulting S-matrix for the scattering of the soliton 
with the triplets of resonances should be solution of the
Yang-Baxter equation corresponding to  the universal $\CR$-matrix
of $\CU_q (\hat{sl(2)} )$ evaluated in the representations 
$V_{1/2} \otimes V_1$ where $V_j $ is the $2j+1$ dimensional
$q$-deformed spin-$j$ representation.

\def\Sn{S^{(n)}}
\def\ha{{1/2}}

Continuing to close the bootstrap leads to resonances in the
q-deformed spin $j$ representation $V_j$ with $j=1,3/2,2,...\infty$. 
To simplify matters we will only consider the scattering of the particle
in $V_j$  of maximal topological charge $+2j$.  This avoids the 
complexities of non-diagonal scattering and will be sufficient 
to obtain a mass formula as a function of $j$. 

Let $\Sn_{\ha,j} (\rap )$ denote the diagonal scattering of the
$n$-th spin $j$ particle of topological charge $2j$ with the 
soliton of topological charge $+1$.  For $j=1$, $n$ refers to the
quantum number in eq.~(\ref{res6}), and  when $j=1/2$, there
is no $n$-dependence and $\Sn_{\ha,\ha} = S_{++}^{++}$.
If the particle of charge $2(j+1/2)$ appears as a resonance
in the scattering of spin $1/2$ with spin $j$ particles, then the
S-matrix has a simple pole:
\beq
\label{st1}
\Sn_{\ha , j} (\rap ) \sim \inv{\rap - \mu_{\ha,j}^{j+1/2} },
\eeq
and the masses are related by the formula
\beq
\label{st2}
m_c^2 = m_a^2 + m_b^2 + 2 m_a m_b \cosh \mu_{ab}^c, 
\eeq
with $a,b,c=1/2,1,3/2,...$.   The above equation implies 
the following relation
between the fusion angles:
\beq
\label{st3}
\mu_{ab}^c + \mu_{bc}^a + \mu_{ac}^b = 2\pi i.
\eeq
In the equations (\ref{st2},\ref{st3}) all the masses are implicitly 
at the same fixed $n$.

\def\mub{\bar{\mu}}

Given $\Sn_{\ha,j}$ one can construct the S-matrix
$\Sn_{\ha,j+\ha}$ from the bootstrap procedure:
\beq
\label{st4}
\Sn_{\ha, j+\ha} (\rap ) = 
\Sn_{\ha, j} \(\rap + \mub_{j,j+\ha}^\ha \) 
\; S_{\ha,\ha} \(\rap - \mub_{\ha,
j+\ha}^j \), 
\eeq
where $\mub = i\pi - \mu$. 

Let us first consider the scattering of the spin $1/2$ solitons with the spin
$1$ resonances discussed above.  We already found that
$\mu_{\ha , \ha}^1 = 2\pi n/\theta $ and eq.~(\ref{st3}) 
implies $\mub_{\ha,1}^\ha = \pi n/\theta$.  The bootstrap
equation then reads 
\beq
\label{st5}
\Sn_{\ha, 1} (\rap ) = S_{\ha,\ha}\(\rap + \frac{n\pi}{\theta} \) 
\; S_{\ha,\ha} \(\rap - \frac{n\pi}{\theta} \).
\eeq
The factor $S_{\ha,\ha} (\rap)$ only has {\it real} poles
at $z= {\beta \theta}/ {2 \pi} = p$ 
and zeros at $z=-p$ where $p=1,2,3,..$.   Remarkably
this leads to the fact that the only real pole of $\Sn_{\ha,1} (\rap )$ is
at $\rap = n\pi/\theta \equiv \mu_{\ha,1}^{3/2} $.  
The other poles of $S_{\ha,\ha}$ either lead to double poles
or are canceled by the zeros in the other $S_{\ha,\ha}$ factor.  

Let $m_n (j)$ denote the mass of the n-th resonance with spin $j$.  
The above value of $\mu_{\ha , 1}^{3/2}$ leads to 
\beq
\label{st6}
m^2_n (j=3/2) = 4M_s^2 \( \frac{9}{4} + 2 \, \sinh^2 \frac{n\pi}{\theta} \). 
\eeq

\def\h{H}

Moving on to the spin $2$ resonances, one finds that the fusion angles
are somewhat more complicated.   Since the masses $m(j)$ are
already determined for $j=1/2,1,3/2$, the relevant fusion angles
follow from eq.~(\ref{st3}) and one obtains
\barray
\nonumber 
\mu_{1/2,1}^{3/2} &=& \frac{n\pi}{\theta},
\\ 
\label{st7}
\exp\mub_{1/2,3/2}^1 &=& \sqrt{ \frac{2 + \h}{2 + \h^{-1}}}, 
\qquad\qquad\qquad\qquad
\h\equiv e^{2 n\pi/\theta},
\\
\nonumber
\exp\mub_{1,3/2}^{1/2}  &=& \sqrt{ \frac{(1+\h)(2+\h^{-1})}{
(1+\h^{-1})(2 + \h)}}.
\earray
The S-matrix is 
\beq
\label{st9}
\Sn_{1/2,3/2} (\rap) = 
\Sn_{1/2,1}\(\rap+\mub_{1,3/2}^{1/2}\) S_{1/2,1/2}\(\rap-\mub_{1/2,3/2}^1\).
\eeq
Similarly to the $j=3/2$ case, the above S-matrix has a 
{\it single} simple pole at $\rap = \mu_{1/2,3/2}^2 $ where
\beq
\label{st10}
\mu_{1/2,3/2}^2 = \frac{n\pi}{\theta} - \mub_{1,3/2}^{1/2}. 
\eeq
This leads to the mass 
\beq
\label{st11}
m_n^2 (j=2) = 4 M_s^2 \( 4 + 3 \sinh^2 \frac{n\pi}{\theta} \). 
\eeq

Having understood the above cases we can readily extend the results to
arbitrary $j$.  Examining the poles in the S-matrix $S_{1/2,j}$ one
finds a single, real, simple pole at 
\beq
\label{st12}
\mu_{1/2,j}^{j+1/2} = \mu_{1/2,j-1/2}^j - \mub_{j-1/2,j}^{1/2}.
\eeq
This leads to the mass formula
\barray
\label{st13}
m^2_n (j) &= & 4 M_s^2 \( j^2 + (2j-1) \sinh^2 \frac{n\pi}{\theta} \) \\
\nonumber  
& =& M_s^2 \, ( 2 j -1 + \h)(2 j -1 + \h^{-1}).
\earray
 From this mass formula one can determine all the fusion angles:
\barray
\nonumber
\exp \mu_{1/2,j}^{j+1/2} &=&
\sqrt{ \frac{2 j -1 + \h}{2 j -1 + \h^{-1}}},
\\ 
\label{st14}
\exp \mub_{1/2,j+1/2}^j &=&
\sqrt{ \frac{2 j  + \h}{2 j  + \h^{-1}}},
\\
\nonumber
\exp \mub_{j,j+1/2}^{1/2} &=&
\sqrt{ \frac{(2j-1 + \h) (2 j  + \h^{-1})}{
(2 j -1  + \h^{-1})(2 j + \h)}}.
\earray

The above mass spectrum is suggestive of a string theory. 
We remark that the higher spin particles that make up the
Regge trajectory of the bosonic string also appear 
as resonances on the physical cut in the Veneziano 
amplitude~\cite{strings}.   Consider
the ``leading Regge trajectory'', i.e.\ the lowest mass at spin $j$. 
When $\theta$ is very small,  the formula~(\ref{st13}) becomes
\beq
\label{st15}
m^2 (j) \approx  \inv{\alpha'} (j - \alpha (0) ),
\eeq
where the Regge slope and intercept are 
\beq
\label{st16}
\inv{\alpha'} = 2 M_s^2 e^{2\pi/\theta} , ~~~~~~\alpha (0) = 1/2.
\eeq
(The leading Regge trajectory corresponds to $n=1$.) 
The usual field theory limit in string theory is $\alpha' \to 0$
which  corresponds to $\theta \to 0$, where all the resonances become
infinitely massive. The fact that our theory has two
coupling constants, namely $M_s$ and $\theta$, also
suggests that the underlying string has another parameter
such as a compactification radius.  
If there is indeed an underlying string description,  much remains
to be understood regarding the nature of this string theory.  

\subsection{Jumping the $c_m=1$ barrier?}

We close this section with some speculative remarks 
regarding the possible nature of the string theory, if there
is indeed such a description.   
Consider a non-critical bosonic string theory in $D$ spacetime
dimensions,  where $c_m = D-1 $ is the
Virasoro central charge of the matter content with $c_m < 26$.  
Because $c_m \neq 26$, there are some Liouville degrees of freedom
with central charge $c_L$,  where $c_m + c_L = 26$.   The Liouville
theory has the action
\beq
\label{liou}
S_{\rm Liouville} = 
\int \frac{d^2 x}{4\pi} \( 
\inv{2} (\d \phi )^2 + \lambda e^{i b\phi}  \). 
\eeq
The central charge of the Liouville theory is related to $b$ as
follows:
\beq
\label{liou2}
c_L = 1 + 6 \( \frac{i b}{\sqrt{2}} + \frac{\sqrt{2}}{i b} \)^2.  
\eeq
The full range $1<c_m<25$ beyond the ``$c_m = 1$'' barrier, 
corresponds to 
\beq
\label{liou5}
b^2 = 2 e^{i\alpha}, ~~~~~{\rm with} ~~~~ \cos \alpha = 
\frac{c_m - 13}{12}
\eeq
Note that in the range we are considering, $-1 < \cos \alpha < 1$. 
Like the sine-Gordon theory,  the Liouville theory has a finite
$\CU_q (sl(2))$ quantum group symmetry with $q$ the same as in 
eq.~(\ref{que})~\cite{GomezS,nonlocal}.

To relate our S-matrix theory to the string theory we would
need to relate the Polyakov path integral over Riemann surfaces
(summed over genus) 
for the matter coupled to 2d gravity, and it isn't at all clear
how to even begin to do this.  We point out however that this
has been done for the $c_m =1$ case leading to an exact 2D S-matrix
\cite{moore}.  (For a review, see\cite{ginsparg,klebanov}.)
Here we merely examine the following. 
The $e^{ib \phi}$ interaction of the Liouville theory 
is one part of the $\cos b\phi$ interaction of the sine-Gordon theory,
and we expect that the Liouville theory for $b$ given
in eq. (\ref{liou5}) shares some common features with the
sine-Gordon theory at the same $b$. 

For $\alpha = -\theta/2$, and $\theta$ small, eq. (\ref{liou5}) 
matches eq. (\ref{bos4}) and thus corresponds to our cyclic
sine-Gordon model.  One sees that this is just below the 
$c_m = 25$ barrier:
\beq
\label{liou6}
c_m \approx 25 - \frac{3}{2} \theta^2 
\eeq
Interestingly, when $\alpha \approx \pi -\theta/2$ with 
$\theta$ small, $c_m$ is just above 1, 
$c_m \approx 1 + 3\theta^2 /2 $,  and this corresponds to 
the deformation of the {\bf sinh}-Gordon theory which is
the staircase model\cite{staircase1}.   
It would be very interesting if the idea of a Russian doll
RG plays a role in resolving the known long-standing difficulties
in going beyond $c_m =1$.

\subsection{Comparison between the cyclic sine-Gordon model
and Russian doll BCS model}

It is interesting to make a comparison between
the cyclic sine-Gordon model we have studied
in the previous sections and the Russian
doll BCS model introduced in reference~\cite{LRS}, 
which also possesses a cyclic RG. 
The model considered in the latter reference
is a simple modification of the reduced
BCS model used in the study of the superconducting properties
of ultrasmall metallic grains. It describes the pairing
interactions between $N/2$ pairs of electrons occupying $N$
energy levels $\varepsilon_j$, which are separated by an 
energy distance $2 \delta$, i.e.\ $\varepsilon_{j+1} - 
\varepsilon_j = 2 \delta$. The model is characterized by 
a scattering potential
$V_{j,j'}$ equal to $g + i \, \theta$ for 
$\varepsilon_j >\varepsilon_{j'}$ and 
  $g - i\,  \theta$ for 
$\varepsilon_j  < \varepsilon_{j'}$
in units of the energy spacing $\delta$. 

The case $\theta =0$ is equivalent to the 
usual BCS model which has a unique
condensate with a superconducting gap
given by $\Delta \sim 2 N \delta e^{-1/g}$
for $g$ sufficiently small. When 
$\theta$ is non zero the BCS gap equation
has an infinite number of solutions
corresponding to condensates with
gaps $\Delta_n(N) \sim N A \delta  e^{-n \pi/\theta}
\;\; (n=0,1,\dots,\infty)$, where $A$ depends
on $g$ and $\theta$. This model can also 
be studied using a renormalization group
which reduces the number of energy levels
$N$, namely $N(s) = e^{-s} N$, 
by integrating out the high energy modes.  
The RG leaves invariant 
the coupling constant 
$\theta$, while $g(s)$ runs with an equation
similar to the one loop result~(\ref{one4}) and has a cyclic
behavior with a period $\lambda_{BCS} = \pi/\theta$,
which in turn is related to the cyclicity
of the gaps 
$\Delta_{n}(e^{- \lambda_{BCS}} N ) \approx \Delta_{n+1}(N)$.
Notice that in this case,  
as compared to eq.~(\ref{res7}),
the jump in $n$ is one,  
due to the fact that 
there is no discrete symmetry
associated to $n$.
In a finite system the number of the 
condensates $n_c$ is given 
by $n_c \sim \frac{\theta}{\pi} \log N$, 
in close resemblance to eq.~(\ref{res8}). 
All these results together with the
corresponding ones for the cyclic sine-Gordon
model are summarized in table~2.

\begin{center}
\begin{tabular}{|c|c|c|}
\hline 
& Cyclic sine-Gordon & Russian doll BCS \\
\hline \hline 
RG-time & $L= e^l \, a $ & $N(s) = e^{-s} \, N $ \\
\hline
RG-period & $\lambda_{CSG} = \frac{2 \pi}{\theta}$ & 
$\lambda_{BCS} = \frac{\pi}{\theta}$ \\
\hline
 & Resonances & Condensates \\
Energy scales
& $m_n(L) \sim \frac{1}{L} e^{n \pi/\theta}$ &
 $\Delta_n(N) \sim N A e^{-n \pi/\theta}$ \\
\hline
Russian doll scaling & 
 $m_n(e^{-\lambda_{CSG}}  L ) \approx   m_{n+2} (L)$ & 
$\Delta_{n}(e^{- \lambda_{BCS}} N ) \approx \Delta_{n+1}(N)$ \\
\hline
Finite systems
& 
$n_{\rm res} \sim \frac{\theta}{\pi} \log(L/a)$ &
  $n_c \sim \frac{\theta}{\pi} \log N$ \\
\hline 
\end{tabular}

\vspace{0.5 cm}
Table 2.- 
Comparison between the cyclic sine-Gordon model
and the Russian doll BCS models.
\end{center}

\section{Conclusions}

In this paper we have analyzed in detail
the cyclic regime of the Kosterlitz-Thouless flows
of a current-current perturbation of the
$su(2)$ level $k=1$ WZW model~\cite{BL}. 
We have computed the RG time $\lambda$ of a complete
cycle in terms of the RG invariant $Q=- \theta^2/16$ of the non perturbative
RG equations~\cite{GLM}, obtaining $\lambda = 2 \pi/\theta$. 
Using standard bosonization techniques we have
mapped the perturbed WZW model into the sine-Gordon
model parametrized by a complex coupling 
$b^2 = 2/(1 + i \theta/2)$. The latter model 
possesses a quantum affine algebra
$\CU_q (\hat{sl(2)} )$ with a real  
quantum deformation parameter $q = e^{- \pi \theta/2}$,
together with a dual quantum
group structure $\CU_{\tilde{q}} (\hat{sl(2)} )$
with $\tilde{q}^{-1}  = e^{2\pi/ \theta} = e^{\lambda}$.
Assuming the existence of solitons with
topological charge $\pm 1 $ with 
a factorized scattering S-matrix commuting with the 
$\CU_q (\hat{sl(2)} )$
 symmetry, we
have 
examined two different S-matrix solutions differing by
overall scalar factors. 

The first S-matrix we considered in section IV is real analytic
and thus strictly unitary in the usual sense.  It can also be
obtained from a certain limit of the XXZ spìn chain.  The
S-matrix possesses a cyclicity as a function of energy,
eq. (\ref{scyclicity}) which is a clear signature of a cyclic
RG.  We conjectured that it describes the cyclic regime of 
the KT flows.

The other S-matrix we considered in section VI   
is an analytic extension of the usual 
massive sine-Gordon S-matrix.  
In this case the  cyclic S-matrix has
poles leading to an infinite number of resonances with the Russian
doll scaling behavior also consistent with a cyclic RG flow.
However since this S-matrix is not unitary in the usual sense
because of the lack of real analyticity, it cannot describe
the hermitian theory defined by the current-current perturbation. 
Thus it's physical meaning remains unclear.  If there is a physically
sensible underlying model, it probably corresponds to a parity
violating theory as discussed in\cite{Mira}.  
 Perhaps such a theory can 
be obtained as a certain limit of the XYZ chain.

Remarkably we found that closure of the cyclic sine-Gordon 
S-matrix bootstrap
leads to resonances of arbitrarily higher spin with a string-like
mass formula.   As discussed in section VI, 
it would be very interesting if the idea of a cyclic RG could
shed new light on the problem of the $c=1$ barrier.  

A few remarks concerning the so-called c-theorem are warranted. 
The c-theorem is a statement regarding the irreversibility of
the RG flow: as the scale is increased massive states decouple
and $c$ decreases monotonically \cite{ctheorem}.  The c-function 
may be viewed as a function of the scale-dependent couplings,
$c(L) = c(g(L) )$.   Thus, a cyclic RG would appear to give rise
to a periodic c-function, violating the c-theorem.  A
thermodynamic Bethe ansatz analysis \cite{tba} of the c-function 
based on the S-matrix  proposed in section IV  indeed reveals 
a periodic c-function in the deep ultraviolet \cite{LRS3}.
How can this violation of the fundamental c-theorem
occur  in a unitary theory as commonplace  as the KT flows? 
One answer is  that Zamolodchikov's proof of the c-theorem assumes
that one can define the theory as a perturbation about a UV fixed
point,  however  a cyclic RG by definition has no such fixed point.

This work further supports the idea that cyclic RG behavior may be
more commonplace than originally anticipated, and may possibly
represent a new paradigm for Physics. It is tempting, though probably
foolhardy,  to  even speculate
that the generations of fermions in the standard model 
are just the first few in an infinite sequence of Russian
dolls.

\section*{Acknowledgments} 

We would like to thank D. Bernard, S. Lukyanov and L. Miramontes
 for discussions.  This work has been supported by the Spanish grants 
SAB2001-0011 (AL), 
BFM2000-1320-C02-01 (JMR and GS), and by the NSF of the USA.

\vfill\eject

\end{document}